\documentclass[showpacs,preprint]{revtex4}
\usepackage{graphicx}
\usepackage{subfigure}
\usepackage{amssymb}
\usepackage{amsmath}
\usepackage{bm}
\usepackage{latexsym}
\usepackage{hyperref}

\begin{document}

\begin{titlepage}
\title
{Non-Markovian coherence dynamics of driven spin boson model: damped
quantum beat or large amplitude coherence oscillation}
\author
{Xiufeng Cao\footnote{Email: cxf@sjtu.edu.cn}, Hang Zheng}
\address{Department of Physics, Shanghai Jiaotong University,
Shanghai 200240, People's Republic of China}

\begin{abstract}
The dynamics of driven spin boson model is studied analytically by
means of the perturbation approach based on a unitary
transformation. We gave the analytical expression for the population
difference and coherence of the two level system. The results show
that in the weak driven case, the population difference present
damped coherent oscillation (single or double frequency) and the
frequencies depend on the initial state. The coherence exhibit
damped oscillation with Rabi frequency. When driven field is strong
enough, the population difference exhibit undamped large-amplitude
coherent oscillation. The results easily return to the two extreme
cases without dissipation or without periodic driven.
\\


\end{abstract}

 \maketitle

\end{titlepage}

1. Introduction

Discrete\ electronic states of the qubits can be considered as localized two
levels systems (TLS), which has been related to the potential use as
building blocks of prospective quantum logic gates \cite{1,2}. Solid-state
qubits, including superconducting \cite{3,4,5,6} and semiconductor\cite%
{7,8,9,10} ones, couple to environmental degrees of freedom that potentially
lead to dephasing. The development of techniques to protect coherence of
carrier tunneling in the qubits are key points for a successful
implementation of quantum information processing in these systems and the
development of qubit based quantum information systems.

Since the development of local spectroscopy techniques, the
atomiclike optical properties of semiconductor quantum dots (charge
qubit) or superconductive ring (flux qubit) have been intensively
studied. The group of T. H. Stievater have reported the first
observation of Rabi oscillations from excitons confined to single
GaAs/Al$_{0.3}$Ga$_{0.7}$As QD\cite{11}. L. Besombes \textit{et al.}
optically control the charge state of a single QD and coherently
manipulate the confined wave function exploiting quantum
interference and Rabi oscillation phenomena by microspectroscopy in
individual InGaAs/GaAs QDs\cite{12}. Using a pulse technique,
Pashkin \textit{et al.} coherently mix quantum states and observe
quantum coherent oscillation of coupled qubits in the vicinity of
the co-resonance\cite{13} , the spectrum of which reflects
interaction between the qubits. Zrenner \textit{et al.} actualize
Rabi oscillation by playing an InGaAs quantum dot in a photodiode
and demonstrate that coherent optical excitations in the quantum dot
two level system can be converted into deterministic
photocurrents\cite{14}. While many-body effects of the environment
fundamentally change many aspects of Rabi oscillations, particularly
the lack of saturation or decoherence. Thus, we think over two
questions: first, how does the interplay or competition between
optical driving and dissipation in the TLS influence the
decoherence? second, can one sustain large-amplitude coherent
oscillation in driven spin boson model? A good knowledge of the
decoherence holds most prominently for many applications such as
optoelectronic devices in quantum information processing where the
operation completely relies on the presence of coherence\cite{15}.

The dynamics of driven spin-boson model\cite{16} also attract
theoretical physics attention for its widespread applications to
various biological, chemical, and physical systems, e.g., ac-driven
superconducting quantum interference devices, laser-induced
isomerization of bistable molecules, laser-induced localization of
electrons in semiconductor double-well quantum structures, or
paraelectric resonance. The various communities typically rely on
different methods of description. The most direct approaches is
portrayal of the time evolution of the corresponding reduced density matrix $%
\rho (t),$ which starts from the generalized master equation (GME) of the
TLS. Two popular approximation are either based on the system-bath coupling
expansion obtained by use of a projector operator method (commonly known as
the Bloch-Redfield formalism\cite{17}), or on the expansion in the coupling
expansion in the coupling matrix element $\Delta $ (such as a tunnel
splitting) by use of (real-time) path-integral methods, such as the
application of the so termed noninteracting blip approximation (NIBA\cite{18}%
). Ludwig Hartmann \textit{et al.} have enlighten the advantages and
disadvantages of two approaches\cite{19}. A special case where the
field frequency is comparable to the TLS frequency (resonance or
near-resonance) is prominent important in experiment, but is more
difficult to handle analytically. If the system-bath coupling is
weak, the traditional optical Bloch equations will produce a
meaningful result\cite{20,21}. However, their model with decoherence
rate as a constant parameter is too simple to make detailed
quantitative predictions here\cite{22}. In general, to obtain a
solution for time-dependent spin-boson problems even numerically is
nontrivial task\cite{23,24}.

In this paper we study the quantum dynamics of driven spin-boson
model, where the driven field is near resonance with the TLS.
Analytical explicit expressions for population difference
$\left\langle \sigma _{z}(t)\right\rangle $ and coherence
$\left\langle \sigma _{x}(t)\right\rangle $ are presented through
perturbation treatment based on a unitary transformation. The result
shows that with increasing amplitude of the driven field, the
population difference transferred from damped quantum oscillation
(single-frequency or double-frequency) to large amplitude undamped
coherent oscillation. Therefore, one can efficiently control quantum
coherent dynamic by optical pulse, induce and maintain
large-amplitude coherent oscillations. The critical condition from
damped to undamped large-amplitude coherent oscillations is given.
The coherence decays to ground state with the Rabi frequency in the
case of weak driven.
We also investigated the dependence of the population difference $%
\left\langle \sigma _{z}(t)\right\rangle $ and coherence
$\left\langle \sigma _{x}(t)\right\rangle $ on the initial state.

The paper is organized as follows: In sec 2 we introduce the model
Hamiltonian for driven spin-boson model and solve it in terms of a
perturbation treatment based on unitary transform. We analyze the population
difference and the coherence in different initialization and give discussion
in sec 3. Finally, the conclusion is given in sec 4.

2. The model and theory

We study the driven spin-boson dynamics, where the two level system is
linearly coupled to a continuum of harmonic oscillators and is driven by
classical microwave field. The system under consideration can be modeled by
the Hamiltonian\cite{18,21}:
\begin{equation}
H(t)=H_{s}+H_{d}(t)+H_{b}+H_{i}
\end{equation}%
with%
\begin{equation}
H_{s}=-\Delta \sigma _{x}/2
\end{equation}%
\noindent
\begin{equation}
H_{d}(t)=\varepsilon (t)\sigma _{z}
\end{equation}%
\begin{equation}
H_{b}=\sum_{k}\omega _{k}b_{k}^{+}b_{k}
\end{equation}%
\noindent
\begin{equation}
H_{i}=\frac{1}{2}\sum_{k}g_{k}(b_{k}^{+}+b_{k})\sigma _{z}
\end{equation}%
where $H_{s}$ is the Hamiltonian of the TLS, $H_{d}(t)$ of the
external driven field, $H_{b}$ of the bath and $H_{i}$ of bath and
system interaction that is responsible for decoherence. Throughout
this paper we set $\hbar =1$. Here \noindent \noindent $\sigma _{i}$
are pauli spin matrices, $\Delta $ describe the coupling between the
two state and $\varepsilon (t)$ is the external time dependent
modulated field. $b_{k}^{+}(b_{k})$ and $\omega _{k}$ are the
creation (annihilation) operator and energy of the phonons with the
wave vector $k$. $g_{k}$ is the electron-phonon coupling strength.
The ohmic
bath is characterized by its spectral density:%
\begin{equation}
J(\omega )=\sum_{k}g_{k}^{_{2}}\delta (\omega -\omega _{k})=2\alpha \omega
\theta (\omega _{c}-\omega )
\end{equation}%
where $\alpha $ is the dimensionless coupling constant and $\theta (x)$ is
the usual step function.

Firstly we diagonalize the Hamiltonian independent of time
$H_{s}+H_{b}+H_{i}$. Here we apply a canonical transformation,
$H^{\prime }=\exp (s)H\exp (-s)$ with the generator\cite{25}:

\begin{equation}
S=\sum_{k}\frac{g_{k}}{2\omega _{k}}\xi _{k}(b_{k}^{+}-b_{k})\sigma _{z}
\end{equation}

Thus we get the Hamiltonian $H^{\prime }$ and decompose it into three parts:

\begin{equation}
H^{^{\prime }}=H_{0}^{^{\prime }}+H_{1}^{^{\prime }}+H_{2}^{^{\prime }}
\end{equation}%
where
\begin{equation}
H_{0}^{^{\prime }}=-\frac{1}{2}\eta \Delta \sigma _{x}+\sum_{k}\omega
_{k}b_{k}^{+}b_{k}-\sum_{k}\frac{g_{k}^{2}}{4\omega _{k}}\xi _{k}(2-\xi
_{k}^{+})
\end{equation}

\bigskip
\begin{equation}
H_{1}^{^{\prime }}=-\frac{1}{2}\sum_{k}g_{k}(1-\xi
_{k}^{+})(b_{k}^{+}+b_{k})\sigma _{x}-i\frac{\eta \Delta }{2}\sigma
_{y}\sum_{k}\frac{g_{k}}{\omega _{k}}\xi _{k}^{+}(b_{k}^{+}-b_{k})
\end{equation}

\bigskip

\begin{eqnarray}
H_{2}^{^{\prime }} &=&-\frac{1}{2}\Delta \sigma _{x}(\cosh (\sum_{k}\frac{%
g_{k}}{\omega _{k}}\xi _{k}^{+}(b_{k}^{+}-b_{k})-\eta )  \notag \\
&&-i\frac{\Delta }{2}\sigma _{y}(\sinh (\sum_{k}\frac{g_{k}}{\omega _{k}}\xi
_{k}(b_{k}^{+}-b_{k}))-\eta \sum_{k}\frac{g_{k}}{\omega _{k}}\xi
_{k}^{+}(b_{k}^{+}-b_{k}))
\end{eqnarray}%
with

\bigskip

\begin{equation}
\eta =\exp (-\sum_{k}\frac{g_{k}^{2}}{2\omega _{k}^{2}}\xi _{k}^{2})
\end{equation}%
\begin{equation}
\xi _{k}=\frac{\omega _{k}}{\omega _{k}+\eta \Delta }
\end{equation}%
Obviously, $H_{0}^{^{\prime }}$ can be solved exactly. We denote the
ground state of $H_{0}^{^{\prime }}$ as $|g\rangle =|s_{1}\rangle
|\{0_{k}\}\rangle
$, and the lowest excited states as $|s_{2}\rangle |\{0_{k}\}\rangle $, $%
|s_{1}\rangle |\{1_{k}\}\rangle $ where $|s_{1}\rangle $ and $|s_{2}\rangle $
are eigenstates of $\sigma _{x}$ ($\sigma _{x}|s_{1}\rangle =|s_{1}\rangle ,$
$\sigma _{x}|s_{2}\rangle =-|s_{2}\rangle $), $|\{n_{k}\}\rangle $ means
that there are $n_{k}$ phonons for mode $k$. The last term of $%
H_{0}^{^{\prime }}$ is constant energy and has no effect to the
dynamics behavior. Thus, we can diagonalize the lowest excited
states $H^{^{\prime
}} $ as:%
\begin{equation}
H^{^{\prime }}=-\frac{1}{2}\eta \Delta |g\rangle \langle
g|+\sum_{E}E|E\rangle \langle E|+\text{terms with higher excited states}
\end{equation}

The digitalization is through the following transformations\cite{25}%
\begin{equation}
|s_{2}\rangle |\{0_{k}\}\rangle =\sum_{E}x(E)|E\rangle
\end{equation}%
\begin{equation}
|s_{1}\rangle |\{1_{k}\}\rangle =\sum_{E}y_{k}(E)|E\rangle
\end{equation}

\begin{equation}
|E\rangle =x(E)|s_{2}\rangle |\{0_{k}\}\rangle
+\sum_{k}y_{k}(E)|s_{1}\rangle |\{1_{k}\}\rangle
\end{equation}%
where
\begin{equation}
x(E)=(1+\sum_{k}\frac{V_{k}^{2}}{(E+\frac{1}{2}\eta \Delta -\omega _{k})^{2}}%
)^{\frac{1}{2}}
\end{equation}%
and%
\begin{equation}
y_{k}(E)=\frac{V_{k}}{E+\frac{1}{2}\eta \Delta -\omega _{k}}x(E)
\end{equation}%
with $V_{k}=\eta \Delta g_{k}\xi _{k}/\omega _{k}.$ E are the
diagonalized excitation energy and they are solutions of the
equation
\begin{equation}
E-\frac{1}{2}\eta \Delta -\sum_{k}\frac{V_{k}^{2}}{E+\frac{1}{2}\eta \Delta
-\omega _{k}}=0
\end{equation}

A series of experiments have successfully realized coherent control of the
qubit by applying resonant microwave excitations\cite{3}. The qubit state
evolves driven by a time-dependent term $\varepsilon _{mw}\cos (2\pi
Ft)\sigma _{z}$ in the Hamiltonian, where F is the microwave frequency and $%
\varepsilon _{mw}$ is the microwave amplitude. So we focus our
attention on a monochromatic field of the form $\varepsilon
(t)=\varepsilon \cos (\Omega t)\sigma _{z}\ $and choose the control
field excitation to be resonant with the splitting of the tunneling
TLS such that it also does not induce transition to higher-lying
excitation. So Hamiltonian can approximately be describes as:

\begin{equation}
H^{^{\prime }}=-\frac{1}{2}\eta \Delta |g\rangle \langle
g|+\sum_{E}E|E\rangle \langle E|-\varepsilon \cos (\Omega t)\sigma _{z}
\end{equation}%
After expand with eigenstate $|g\rangle $ and $|E\rangle :$
\begin{equation}
H^{^{\prime }}=-\frac{1}{2}\eta \Delta |g\rangle \langle
g|+\sum_{E}E|E\rangle \langle E|-\frac{\varepsilon }{2}\sum_{E}(x^{\ast
}(E)|g\rangle \langle E|\exp (i\Omega t)+x(E)|E\rangle \langle g|\exp
(-i\Omega t))
\end{equation}

In deriving Eq.22 we have ignored the nonenergy conserving term or
counter rotation term in the rotation wave approxmation\cite{26}.
This is generally a very good approximation, especially in the
special case when the two states are at resonance or near resonance
with the incident field $\Delta \approx \Omega .$ In the interaction
picture,

\begin{equation}
H_{0I}^{^{\prime }}=-\frac{1}{2}\eta \Delta |g\rangle \langle
g|+\sum_{E}E|E\rangle \langle E|
\end{equation}

\begin{eqnarray}
V_{I}(t) &=&-\frac{\varepsilon }{2}\sum_{E}(x^{\ast }(E)|g\rangle \langle
E|\exp (i(-\frac{1}{2}\eta \Delta -E+\Omega )t)  \notag \\
&&+x(E)|E\rangle \langle g|\exp (-i(-\frac{1}{2}\eta \Delta -E+\Omega )t))
\end{eqnarray}

The wave function can be written in the form $|\Psi (t)\rangle
=C_{1}(t)|g\rangle +\sum_{E}C_{E}(t)|E\rangle ,$ where $C_{1}(t)$ and $%
C_{E}(t)$ are the probability of find the particle in state
$|g\rangle $ and $|E\rangle $ at time t, respectively. The
corresponding Heisenberg equation is
\begin{equation}
\frac{d}{dt}|\Psi (t)\rangle =-iV_{I}(t)|\Psi (t)\rangle
\end{equation}

Assumed that the initial state of the model is $\exp (S)|\Psi
(0)\rangle =C_{1}(0)|s_{1}\rangle |\{0_{k}\}\rangle
+C_{2}(0)|s_{2}\rangle |\{0_{k}\}\rangle =C_{1}(0)|g\rangle
+C_{2}(0)\sum_{E}x(E)|E\rangle $. Using Laplace transformation, we
obtain

\begin{equation}
C_{0}(P)=\frac{C_{1}(0)+C_{2}(0)\frac{\varepsilon }{2}\sum_{E}\frac{\mid
x(E)\mid ^{2}}{E-iP}}{P+i(-\frac{1}{2}\eta \Delta +\Omega )-i\frac{%
\varepsilon ^{2}}{4}\sum_{E}\frac{\mid x(E)\mid ^{2}}{E-iP}}
\end{equation}%
and%
\begin{equation}
\sum_{E}x^{\ast }(E)C_{E}(P)=\frac{(C_{2}(0)\ast (P+i(-\frac{1}{2}\eta
\Delta +\Omega ))+iC_{1}(0)\frac{\varepsilon }{2})\ast -i\sum_{E}\frac{\mid
x(E)\mid ^{2}}{E-iP}}{P+i(-\frac{1}{2}\eta \Delta +\Omega )-i\frac{%
\varepsilon ^{2}}{4}\sum_{E}\frac{\mid x(E)\mid ^{2}}{E-iP}}.
\end{equation}%
In the complex function theory, the sum of E can be simplified:%
\begin{eqnarray}
\sum_{E}\frac{\mid x(E)\mid ^{2}}{E-iP} &=&\frac{1}{2\pi i}\int \frac{dE}{(E-%
\frac{1}{2}\eta \Delta -\sum_{k}\frac{V_{k}^{2}}{E+\frac{1}{2}\eta \Delta
-\omega _{k}})(E-iP)} \\
&=&-\frac{1}{iP-\frac{1}{2}\eta \Delta -\sum_{k}\frac{V_{k}^{2}}{iP+\frac{1}{%
2}\eta \Delta -\omega _{k}}}
\end{eqnarray}%
After Changing $iP+\frac{1}{2}\eta \Delta $ to $\omega +i0^{+}$\cite{27}$,$ $%
C_{0}(P)$ and $\sum_{E}x^{\ast }(E)C_{E}(P)$ can be rewritten to:%
\begin{equation}
C_{0}(\omega )=\frac{i(\omega -\eta \Delta -R(\omega )+i\gamma (\omega
))C_{1}(0)-iC_{2}(0)\frac{\varepsilon }{2}}{(\omega -\Omega )(\omega -\eta
\Delta -R(\omega )+i\gamma (\omega ))-\frac{\varepsilon ^{2}}{4}}
\end{equation}%
and%
\begin{equation}
\sum_{E}x^{\ast }(\omega )C_{E}(\omega )=\frac{i(\omega -\Omega
)C_{2}(0)-iC_{1}(0)\frac{\varepsilon }{2}}{(\omega -\Omega )(\omega -\eta
\Delta -R(\omega )+i\gamma (\omega ))-\frac{\varepsilon ^{2}}{4}}
\end{equation}%
where $R(\omega )$ and $\gamma (\omega )$ denote the real and imaginary
parts of $\sum_{k}V_{k}^{2}/(\omega -\omega _{k}),$

\begin{equation}
R(\omega )=-2\alpha \frac{(\eta \Delta )^{2}}{\omega +\eta \Delta }\left\{
\frac{\omega _{c}}{\omega _{c}+\eta \Delta }-\frac{\omega }{\omega +\eta
\Delta }\ln \left[ \frac{\left\vert \omega \right\vert (\omega _{c}+\eta
\Delta )}{\eta \Delta (\omega _{c}-\omega )}\right] \right\}
\end{equation}%
and

\begin{equation}
\gamma (\omega )=2\alpha \pi \omega \frac{(\eta \Delta )^{2}}{(\omega +\eta
\Delta )^{2}}(\theta (\omega )+\theta (\omega _{c}-\omega )-1)
\end{equation}%
Comparing with Bloch equation and Markovian approximation,
decoherence rates $\gamma (\omega )$ become frequency dependent
themselves. $R(\omega )$ is determined by the whole of the spectrum
of frequencies in the spectral density $J(\omega )$ of the ohmic
bath. According to the correlation of real and imaginary parts in
complex function theory, $\gamma (\omega )$ also
involve the whole of the spectrum of frequencies in the spectral density $%
J(\omega )$ of the ohmic bath. That is more general and physical.

Then we inverse Laplace transformation to time parameter space:
\begin{equation}
C_{0}(t)=\frac{\exp (i\eta \Delta t/2)}{2\pi }\int\limits_{-\infty
}^{+\infty }C_{0}(\omega )\exp (-i\omega t+0^{+})d\omega
\end{equation}%
and%
\begin{equation}
\sum_{E}x^{\ast }(\omega )C_{E}(t)=\frac{\exp (i\eta \Delta t/2)}{2\pi }%
\int\limits_{-\infty }^{+\infty }\sum_{E}x^{\ast }(\omega )C_{E}(\omega
)\exp (-i\omega t+0^{+})d\omega .
\end{equation}%
The expectation value $\sigma _{i}$ can be expressed as:%
\begin{equation}
\left\langle \sigma _{i}(t)\right\rangle =\langle \Psi (0)|\sigma
_{i}(t)|\Psi (0)\rangle =\langle \Psi (0)|\exp (-S)\exp (-iH^{^{\prime
}}t)\sigma _{i}\exp (iH^{^{\prime }}t)\exp (S)|\Psi (0)\rangle
\end{equation}%
After straightforward calculation, the electron population difference $%
\left\langle \sigma _{z}(t)\right\rangle $ and the coherence $\left\langle
\sigma _{x}(t)\right\rangle $ are gained as follows:%
\begin{equation}
\left\langle \sigma _{z}(t)\right\rangle =2Re(C_{0}^{\ast
}(t)\sum_{E}x^{\ast }(\omega )C_{E}(t)\exp (-i\Omega t))
\end{equation}%
and
\begin{equation}
\left\langle \sigma _{x}(t)\right\rangle ==\eta \ast (1-2\sum_{E}x^{\ast
}(\omega )C_{E}(t)\sum_{E^{^{\prime }}}x^{\ast }(\omega )C_{E^{^{\prime
}}}(t)).
\end{equation}%
Rather simple expression for the population and coherence are
obtained analytically.

3. The result and discussion

Here we only discuss the near resonant case $\Omega =\Delta .$ $\omega _{c}$
is taken as the energy unit. Without special indication, the coupling
constant of the environment and TLS is taken as $\alpha =0.01.$

In the especial case without coupling to the environment, we rotate
$\sigma
_{z}$ axis around $\sigma _{y}$ axis to $\sigma _{x}$ axis, correspondingly $%
\sigma _{x}$ axis to $\sigma _{z}$ axis, the Hamiltonian turn into
the famous quantum optics model describing the interaction of a
single-mode radiation
with a two-level atom\cite{26}. As we have known, the dynamics of $%
\left\langle \sigma _{z}(t)\right\rangle $ depend on the initial condition.
If the qubit is at $\left\langle \sigma _{x}(t=0)\right\rangle =1$
eigenvalue, the driven field is turned on, then, using the rotating wave
approximation, i.e. in the near resonance condition, the population
difference $\left\langle \sigma _{z}(t)\right\rangle $ will exhibit quantum
beats that result from the interference of fast oscillation with field
frequency $\Omega $ and slow oscillations with Rabi frequency $\varepsilon $%
. The coherence $\left\langle \sigma _{x}(t)\right\rangle $ oscillates with
the Rabi frequency. But if the system is initially in the state $%
\left\langle \sigma _{z}(t=0)\right\rangle =1,$ without the environment, $%
\left\langle \sigma _{z}(t)\right\rangle $ will present $\cos \Omega
t$ oscillation and $\left\langle \sigma _{x}(t)\right\rangle =0$. We
also knew that in the other especial case lack of the driven field,
$\left\langle \sigma _{z}(t)\right\rangle $ damply oscillate
\cite{25}. The question is how will this dissipation environment and
driven field collectively influence the population difference and
coherence of TLS.

In Fig. 1, we plot coherent dynamics of the population difference $%
\left\langle \sigma _{z}(t)\right\rangle $ in the initial state $%
\left\langle \sigma _{z}(t=0)\right\rangle =1,$ where a driving
field is of the same frequency $\Omega =\Delta $, but of different
amplitude. The curves shown in panels (a), (b) and (c) correspond to
three external field amplitude $\varepsilon =0,$ $0.01,$ $0.1$. Fig.
1(d) present population difference $\left\langle \sigma
_{z}(t)\right\rangle $ with $\varepsilon =0.4 $ and $\alpha =0.3.$
The Fourier spectrums of the population difference are presented in
the right four corresponding left panels of Fig. 1. As seen from
Fig. 1(a), in the absence of driven, population difference exhibits
damped oscillation. The frequency $\omega _{0}$ and damping rate
$\gamma $ of the oscillations is well agree with prevenient
result\cite{25}. With weak driving field strength, $\left\langle
\sigma _{z}(t)\right\rangle $ decays with beat pattern in Fig. 1(b)
of two frequency almost $\Omega -\varepsilon $ and $\Omega $.
Further increasing driving field strength over the regime of
half-width $\gamma $ of the driving frequency $\Omega $,
$\left\langle \sigma _{z}(t)\right\rangle $ damped oscillate with a
master frequency $\Omega $ in Fig. 1(c). Above three case shows that
in the weak driven case, the system is always damped and this
genuine quantum coherence oscillation is weakened by friction from
the dissipation environment. For sufficiently strong amplitude
$\varepsilon \geq 2\Omega $, population difference damped a little
at initial short time, then present undamped large-amplitude
coherent oscillations with the frequency of the incident field
$\Omega $. That is to say, the driven field overcome dissipation and
domain the dramatics of the system. It is clearly seen that quantum
beat patter of the population difference $\left\langle \sigma
_{z}(t)\right\rangle $ oscillate only appears in the small regime of
driving strength, which is depend on the frequency $\Omega
-\varepsilon $ and/or $\Omega +\varepsilon $ whether moves out of
the spectral width $\gamma $ of the incident field $\Omega $ or not.

In Fig. 2, the expectation value $\left\langle \sigma
_{x}(t)\right\rangle $ plotted as a function of $\omega _{c}t$ for
TLS driven by a monochromatic field with the same parameters as in
Fig. 1. As seen, in the absence of driving, the coherence
$\left\langle \sigma _{x}(t)\right\rangle $ reaches asymptotically a
value 1, implying that mainly the ground state is occupied at long
times\cite{28}. For weak driving amplitude $\varepsilon =0.01$ and
0.1, it is visible that $\left\langle \sigma _{x}(t)\right\rangle $
oscillated tends to 1 and the oscillation frequency is the Rabi
frequency, which is directly proportional to driven amplitude. While
for large driven field strength $\varepsilon =0.4$, the coherence
$\left\langle \sigma _{x}(t)\right\rangle $ oscillates around zero
with a small amplitude. Driven field make the transfer probability
to left state and right state equal,
that is to say the electron tunnel between the two states left dot state $%
\left\vert L\right\rangle $ and right state $\left\vert R\right\rangle ,$ so
the undamped oscillation occurs as seen in Fig. 1(d). The situation of
strong driving field can be correspond to the case of absence of damping,
where the coherence $\left\langle \sigma _{x}(t)\right\rangle $ is exactly
equal to zero.

Since any superposition $\left\vert \Psi \right\rangle =\alpha \left\vert
s1\right\rangle +\beta \left\vert s2\right\rangle $ can be prepared in
experiment, through manipulation of the quantum state is performed by
applying microwave pulses $\mu (t)$ to the gate\cite{29}. Fig. 3 gives the
population difference $\left\langle \sigma _{z}(t)\right\rangle $ for the
same parameters as in Fig. 1, but assume that at time $t=0$ the particle is
held at superposition $\left\langle \sigma _{x}(t=0)\right\rangle =1$, with
the bath being in thermal equilibrium. Without driven field, $\left\langle
\sigma _{z}(t)\right\rangle =0$. For weak driven, despite the dissipation we
were able to induce Rabi oscillations by applying microwave pulses at the
near-resonance frequency of the TLS. The oscillation decay nonexponentially
and display a clear beating. From the Fourier spectrum, we find two
oscillation frequency, one is the frequency of the incident field $\Omega
-\varepsilon $ and the other is the Rabi frequency $\Omega +\varepsilon $,
that is to say, $\left\langle \sigma _{z}(t)\right\rangle $ can be fitted by
$\sin \varepsilon t\ast \sin \Omega t.$ This corresponds to the common
quantum beat in the quantum optics. The frequencies of population difference
oscillation are dependent on the initial state.\ In Fig.3(d), a well-behaved
coherent oscillations can be observed after adequate long time when the
driven field strength is adequately strong $\varepsilon =0.4$. The frequency
of the oscillations firstly is $\Omega +\varepsilon $, then $\Omega .$

In the same parameters and the initialization with Fig. 3, coherence $%
\left\langle \sigma _{x}(t)\right\rangle $ are plotted in Fig. 4. In the
presence of tunneling with a dissipation bath and weak driven, due to the
dominance of transitions to the continuum modes, we see that the coherence $%
\left\langle \sigma _{x}(t)\right\rangle $ presents damped oscillation and
finally stabilizes the ground state\ as seem from the Fig. 4(a) and 4(b).
The frequency of the oscillations is Rabi frequency. The damped can be look
as a leakage of energy\cite{30}. The leakage can be suppressed by enhancing
of the external field. For strong driving field, the long time limit is
shift to nonzero by external field.

The population difference $\left\langle \sigma _{z}(t)\right\rangle
$ will exhibit damping quantum beat, which have be observed in
superconducting qubit experiment\cite{3}. Comparing Fig. 1 and Fig.
3, these curves show clearly that the initialization of TLS
determine the features of the quantum beat. The condition to present
beat in the initial state of Fig. 1 is more rigour than that of Fig.
3. The quantum beat only emerge in a small regime of the driving
field amplitude, that is to say the driving field amplitude
must be in the range of the damping rate $\gamma $ of the driven frequency $%
\Omega $. While in the initialization of Fig. 3, the quantum beat
can always exist, except the driven amplitude enough strong to
undamped oscillation. The frequency of the beat pattern $\Omega $
and $\Omega -\varepsilon $ or/and $\Omega +\varepsilon $ in Fig. 1,
While in Fig. 3, the frequency of the beat pattern $\Omega
-\varepsilon $ and $\Omega +\varepsilon $. It is helpful to
experimenter for the empirical function choose. The difference of
frequency in Fig.1 and Fig.3 is understood qualitatively as follows.
In
the initialization $\left\langle \sigma _{z}(t=0)\right\rangle =1,$ $%
H_{f}(t)=\varepsilon (t)\sigma _{z}$ is the maximum of $H_{f}(t)$ but $H_{i}=%
\frac{1}{2}\sum_{k}g_{k}(b_{k}^{+}+b_{k})\sigma _{z}$ is zero due to $%
|\{n_{k}\}\rangle =|\{0_{k}\}\rangle ,$ so the driven field firstly domain
the oscillation of TLS, then environment and driven field interplay. At the
end the frequency of the oscillation give priority to $\Omega $, only when $%
\Omega -\varepsilon $ or/and $\Omega +\varepsilon $ in the range of
decoherence rate $\gamma $ of the driving frequency $\Omega ,$ quantum beat
emerges$.$ While in the initialization $\left\langle \sigma
_{x}(t=0)\right\rangle =1,$ at the very start, $H_{f}(t)=\varepsilon
(t)\sigma _{z}$ and $H_{i}=\frac{1}{2}\sum_{k}g_{k}(b_{k}^{+}+b_{k})\sigma
_{z}$ are zero together, driving and bath interact so the frequency of the
oscillation are $\Omega -\varepsilon $ and $\Omega +\varepsilon .$

In what follows, we give the transform condition from damped to
undamped oscillation. From derivation,\ we can see that much of the
behavior of the population difference $\left\langle \sigma
_{z}(t)\right\rangle $ and coherence $\left\langle \sigma
_{x}(t)\right\rangle $ can be learned from a study of the
singularities of $C_{0}(\omega )$ and $\sum_{E}x^{\ast }(\omega
)C_{E}(\omega ).$ According the complex-function theory, we analyze
the real part of the energy denominators, ReD($\omega $)$=(\omega
-\Omega )(\omega
-\eta \Delta -R(\omega ))-\frac{\varepsilon ^{2}}{4}$, that is symmetry in $%
\varepsilon .$ Fig. 5 shows ReD($\omega $) with different driving
field amplitude $\varepsilon =0$ (solid line), 0.1 (dashed), 0.3
(dotted), 0.5 (dash-dotted). When $\varepsilon =0,$ the
singularities of $C_{0}(\omega )$ and $\sum_{E}x^{\ast }(\omega
)C_{E}(\omega )$ are $\omega =0$ and $\omega =\omega _{0},$ which
are correspond to zero drive and Lamb Shift due to the dissipation.
$\omega _{0}$ is determined by the function $\omega -\eta \Delta
-R(\omega )=0$\cite{25}. Owing to the driving field, two
singularities are shift. Define the two singularities $\omega _{1}$
and $\omega _{2}$ and assume $\omega _{1}<\omega _{2}$. We find that
two positive singularities are in two sides of $\omega _{0}$ in the
case of small external driving field
amplitude $\varepsilon .$ With the increasing of $\varepsilon ,$ one pole $%
\omega _{1}$ becomes larger and the other $\omega _{2}$ becomes
smaller, at the end the small singularity come to the negative axis.
Furthermore, the imaginary parts of the denominator $\gamma (\omega
)(\omega -\Omega )$ can be understood as a damping rate. Commonly,
they depend on the frequency. Because of the interactions with the
continuous spectral dissipation bath and driving field, the energy
of the localized state is altered to new value $\omega _{1}$ and
$\omega _{2}.$ The nature of the solution depends critically on
whether the energy $\omega _{1}$ and $\omega _{2}$ are within
the band of state $\omega _{k}.$ Therefore, the continuous band of state $%
\omega _{k}$ is confined to the range $0<\omega _{k}<\omega _{c}$,
so the character of solution determined whether $\omega _{1}$ and
$\omega _{2}$ are also within this range. If the values are within
the range $0<\omega _{1},\omega _{2}<\omega _{c}$, the solution has
an important property: dissipation domain the behavior in the
system, after adequate time, the coherence oscillation is damped
out. The reason is that generally imaginary component of denominator
is not zero throughout the continuum band of photon, the decay rate
$\gamma _{1}(\omega )$ and $\gamma _{1}(\omega )$ corresponding to
$\omega _{1}$ and $\omega _{2}$ are finite. That is to say,
the damping rate $\gamma (\omega )$ is nonzero if only in the section of $%
0<\omega <\omega _{c},$ which is the key point of the result. Of
course, if one of the two energy $\omega _{1}$ or $\omega _{2}$ is
outside of the band of continuum state, i.e. $\omega _{1}$ is
negative or $\omega _{2}>\omega _{c}
$, then the corresponding damping rate of the correspond energy vanishes. $%
\left\langle \sigma _{z}(t)\right\rangle $ can't decay to zero and
always exhibits large amplitude coherent oscillation. The condition
damped oscillation or not depend on two roots of ReD($\omega $) in
the dissipation regime or not. Approximatively estimate when
$\varepsilon \geq 2\Omega ,$ population difference can maintain
undamped long coherence oscillation. We illuminate the possibility
to induce and maintain large amplitude coherent oscillation by
applying a resonant control field and present the condition of the
long time coherence. It is favorable for quantum computation. The
result is similar with Ref. 20.\qquad

4. Conclusion

In this paper, we have investigate the quantum optical control
dynamics of the driven spin boson model by a perturbation treatment
based on a unitary transformation. The population difference and
coherence are obtained explicitly. Our approach is not restricted by
the form of spectral density and the initialization. Additionally, a
simple expression for the real part of the energy denominators,
$(\omega -\Omega )(\omega -\eta \Delta -R(\omega
))-\frac{\varepsilon ^{2}}{4},$ allows us to analyze the transform
condition from damped quantum oscillation to undamped oscillation.
We find that, for weak field, the population difference
$\left\langle \sigma _{z}(t)\right\rangle $ presents damping quantum
beat, while for strong field $\varepsilon \geq 2\Omega $,
$\left\langle \sigma _{z}(t)\right\rangle $ preserve large amplitude
coherent oscillation with the frequency $\Omega $. The coherence
$\left\langle \sigma _{x}(t)\right\rangle $ is also studied.
Finally, we hope that this work will stimulate more experimental and
theoretical works in quantum information and computation for quantum
optical control.

Acknowledgments: This work was supported by the China National Natural
Science Foundation (Grants No. 10474062 and No. 90503007).

--------------------

\newpage

\begin{description}
\item {\large FIGURES}

%\begin{figure}
\vspace{0.3cm} %\centerline{\psfig{figure=fig1.eps,width=8.5cm}}
\vspace{0.3cm} %\caption

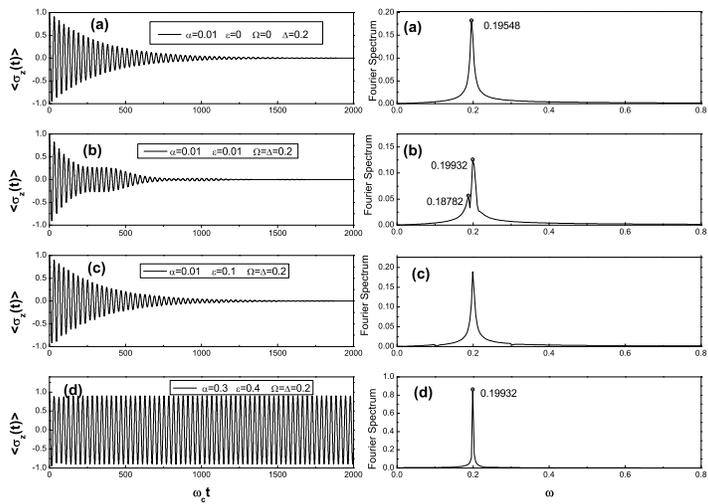
\item {Fig.1. The population difference }$\left\langle \sigma
_{z}(t)\right\rangle ${\ versus }$\omega _{c}t$ in the initialization $%
\left\langle \sigma _{z}(t=0)\right\rangle =1$ with various values of the
driving field amplitude, $\varepsilon =0$, $0.01$, 0.1, 0.4, correspond to
left panels (a) to (d). Also shown with the right panels are the Fourier
spectrum with parameters same as the left panels. the coupling constant of
the environment and TLS $\alpha =0.01$ except (d) $\alpha =0.3.$ The driving
field frequency near-resonance with TLS $\Omega =\Delta .$

%\label{fig1}
%\end{figure}

%\begin{figure}
\vspace{0.3cm} %\centerline{\psfig{figure=fig2.eps,width=8.5cm}}
\vspace{0.3cm} %\caption

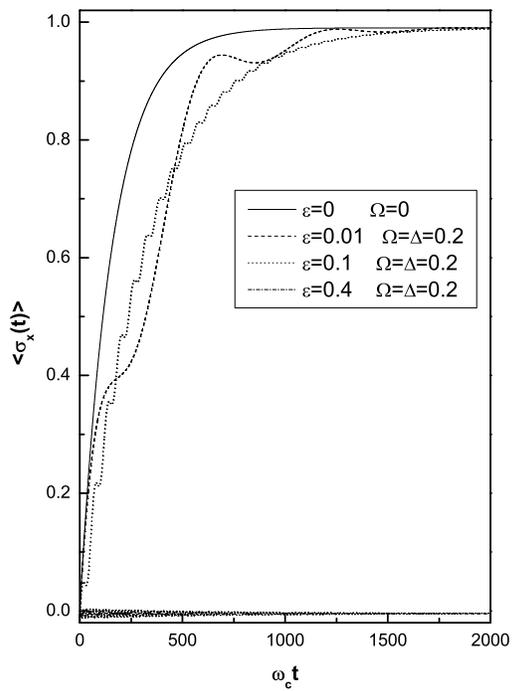
\item {Fig.2. The coherence }$\left\langle \sigma _{x}(t)\right\rangle $ {%
versus }$\omega _{c}t$ in the initialization $\left\langle \sigma
_{z}(t=0)\right\rangle =1$ with various values of the driving field
amplitude, $\varepsilon =0$ (solid line), $0.01$ (dashed line), $0.1$
(dotted line), $0.4$ (dot-dashed line).

%\label{fig1}
%\end{figure}

%\begin{figure}
\vspace{0.3cm} %\centerline{\psfig{figure=fig2.eps,width=8.5cm}}
\vspace{0.3cm} %\caption

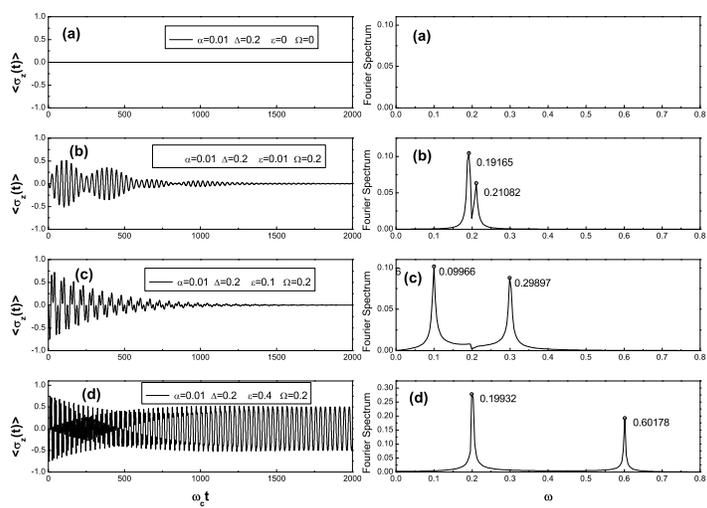
\item {Fig.3. The population difference versus }$\omega _{c}t$ in the
initialization $\left\langle \sigma _{x}(t=0)\right\rangle =1$ with various
values of the driving field amplitude, $\varepsilon =0$, $0.01$, 0.1, 0.4,
correspond to left panels (a) to (d). Also shown with the right panels are
the Fourier spectrum with parameters same as the left panels. the coupling
constant of the environment and TLS $\alpha =0.01.$ The driving field
frequency near-resonance with TLS $\Omega =\Delta .$

%\label{fig1}
%\end{figure}

%\begin{figure}
\vspace{0.3cm} %\centerline{\psfig{figure=fig2.eps,width=8.5cm}}
\vspace{0.3cm} %\caption

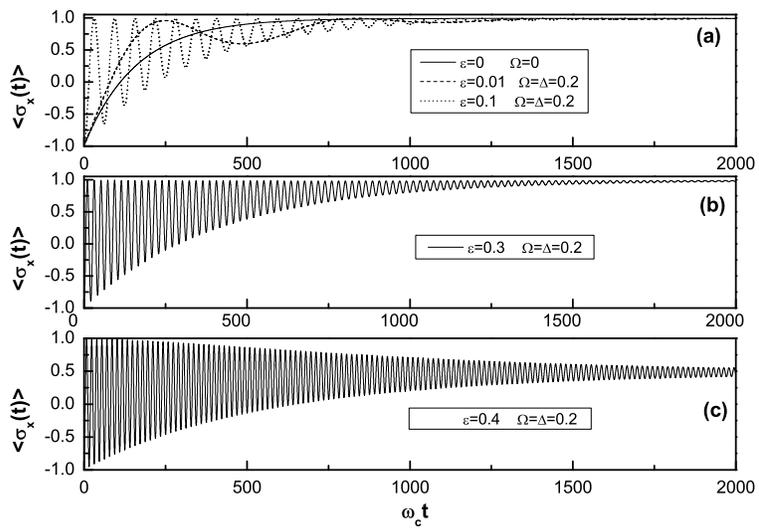
\item {Fig.4} {\ The coherence }$\left\langle \sigma _{x}(t)\right\rangle $ {%
versus }$\omega _{c}t$ in the initialization $\left\langle \sigma
_{x}(t=0)\right\rangle =1$ with various values of the driving field
amplitude. (a) $\varepsilon =0$ (solid line), $0.01$ (dashed line),
$0.1$ (dotted line), (b) $0.3$, (c) $0.4$.

%\label{fig1}
%\end{figure}

%\begin{figure}
\vspace{0.3cm} %\centerline{\psfig{figure=fig2.eps,width=8.5cm}}
\vspace{0.3cm} %\caption

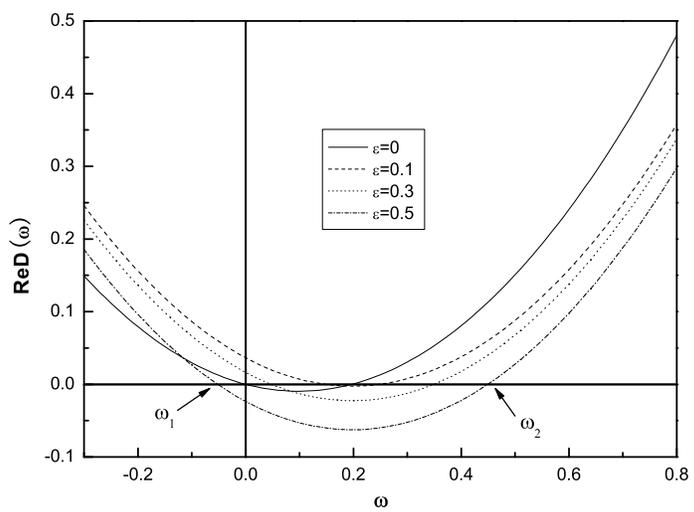
\item {Fig.5} {\ ReD(}$\omega ${)} vs $\omega $ for different driving field
amplitude $\varepsilon =0$(solid line), $0.1$ (dashed line), $0.3$ (dotted
line), $0.5$ (dot-dashed line).

%\label{fig1}
%\end{figure}

%\begin{figure}
\vspace{0.3cm} %\centerline{\psfig{figure=fig2.eps,width=8.5cm}}
\vspace{0.3cm} %\caption
\end{description}

\end{document}